\documentstyle[12pt]{article}
\renewcommand
\baselinestretch{1.4}

\begin{document}

\begin{center}

{\large \bf ZETA-FUNCTION REGULARIZATION IS UNIQUELY DEFINED
AND WELL}

\vspace{4mm}

\renewcommand
\baselinestretch{0.8}
\medskip

{\sc E. Elizalde}\footnote{On leave of absence from and
permanent address: Department E.C.M., Faculty of Physics,
Barcelona University, Diagonal 647, 08028 Barcelona, Spain;
e-mail: eli @ ebubecm1.bitnet}
\\ {\it Division of Applied Mechanics, Norwegian Institute of
Technology, \\
University of Trondheim, N-7034 Trondheim, Norway}

\renewcommand
\baselinestretch{1.4}

\vspace{5mm}

{\bf Abstract}

\end{center}

Hawking's zeta function regularization procedure is shown to
be rigorously and uniquely defined, thus putting and end to
the spreading lore about different difficulties associated
with it. Basic misconceptions, misunderstandings and errors
which keep appearing in important scientific journals when
dealing  with this beautiful regularization method ---and
other analytical procedures--- are clarified and corrected.

\vspace{3cm}


\newpage

This letter is a defense of Hawking's zeta function
regularization method [1], against the different criticisms
which have been published in important scientific journals and
which seem to conclude (sometimes with exagerated emphasis)
that the procedure is ambiguous, ill defined, and that it
possesses even more problems than the well known ones which
afflict e.g. dimensional regularization.
Our main purpose is to clarify, once and for all, some basic
concepts, misunderstandings, and also errors which keep
appearing in the physical literature, about this method of
zeta function regularization and other analytical
regularization procedures. The situation is such that, what is
in fact a most elegant, well defined, and unique ---in many
respects--- regularization method, may look now to the
non-specialist
as just one more among many possible regularization
procedures, plagued with difficulties and illdefiniteness.

We shall not review here in detail the essentials of
the method, nor give an exhaustive list of the papers that
have pointed out `difficulties with the zeta function
procedure'.
For this purpose we refer the specialized reader to a book
scheduled to appear
towards the end of 1993 [2]. Instead, we shall restrict
ourselves to some specific points which are at the very heart
of the matter and which may be interesting to a much broader
audience.

The method of zeta function regularization is uniquely defined
in the following way. Take the Hamiltonian, $H$, corresponding
to our quantum system, plus boundary conditions, plus possible
background field and including a possibly non-trivial metric
(because we may live in a curved spacetime). In mathematical
terms, all this boils down to a (second order, elliptic)
differential operator, $A$, plus corresponding boundary
conditions. The spectrum of this operator $A$ may or may not
be calculable explicitly, and in the first case may or may not
exhibit a beautiful regularity in terms of powers of natural
numbers. Whatever be the situation, it is a well
stablished mathematical theorem that to {\it any} such
operator a {\it zeta function}, $\zeta_A$, can be defined in a
rigorous way. The formal expression of this definition is:
\begin{equation}
\zeta_A (s) = \mbox{Tr}\,  e^{-s \ln A}.
\end{equation}
Let us stress that this is completely {\it general}. Moreover,
the present is by no means the only kind of zeta function
known to the mathematicians, for whom this concept has a wider
character (see, for instance [3]). The zeta function
$\zeta_A (s)$ is generally a meromorphic function (develops
only poles) in the complex plane, $s \in $ C. Its calculation
usually requires
complex integration around some circuit in the complex plane,
the use of the Mellin transform, and the calculation of
invariants of the spacetime metric (the
Hadamard-Minakshisundaram-Pleyel-Seeley-DeWitt-Gilkey-...
coefficients) [4].

In the particular case when the eigenvalues of the
differential operator $A$ ---what is equivalent, the
eigenvalues of the Hamiltonian  (with the boundary conditions,
etc. taken into acount)--- can be calculated explicitly (let
us call them $\lambda_n$ and assume they form a discrete set),
the expression of the zeta function is given by:
\begin{equation}
\zeta_A (s) = \sum_n \lambda_n^{-s}.
\end{equation}
Now, as a particular case of this (already particular) case,
when the
eigenvalues are of one of the forms: (i) $an$, (ii) $a(n+b)$
or (iii) $a(n^2+b^2)$, we
obtain, respectively,  the so-called (i) (ordinary) Riemann
zeta function $\zeta_R$ (or simply $\zeta$), (ii) Hurwitz (or
generalized Riemann) zeta function $\zeta_H$, and (iii) (a
specific
case of the) Epstein-Hurwitz zeta function $\zeta_{EH}$ [3].
Finally, depending on the physical magnitude to be calculated,
the zeta function must be evaluated at a certain particular
value of $s$. For instance, if we are interested in the vacuum
or Casimir energy, which is simply given as the sum over the
spectrum:
\begin{equation}
E_C= \frac{\hbar}{2} \sum_n \lambda_n,
\label{ce}
\end{equation}
this will be given by the corresponding zeta function
evaluated at $s=-1$:
\begin{equation}
E_C= \frac{\hbar}{2} \, \zeta_A (-1).
\end{equation}
Normally, the series (\ref{ce}) will be divergent, and this
will involve an analytic continuation through the zeta
function. That is why such regularization can be termed
as a particular case of analytic continuation procedure.
 In summary, the zeta function of the quantum system is a very
general, uniquely defined, rigorous mathematical concept,
which does not admit either interpretations nor ambiguities.

Let us now come down to the concrete situations which
have motivated this article. In Ref. [5], when calculating the
Casimir energy of a piecewise uniform closed string, Brevik
and Nielsen where confronted with the following expression
([5], Eq. (52))
\begin{equation}
\sum_{n=0}^\infty (n+\beta) ,
\end{equation}
which is clearly infinite. Here, the zeta-function
regularization
procedure consists in the following. This expressions comes
about as the sum over the eigenvalues $n+\beta$ of the
Hamiltonian of a certain quantum system (here the transverse
oscillations of the mentioned string), i.e. $\lambda_n
=n+\beta$. There is little doubt about what to do: as clearly
stated above, the corresponding zeta function is
\begin{equation}
\zeta_A (s) =\sum_{n=0}^\infty (n+\beta)^{-s}.
\end{equation}
Now, for Re $s >1$ this is the expression of the Hurwitz zeta
function $\zeta_H (s;\beta)$, which can be analytically
continued as a meromorphic function to the whole complex
plane. Thus, the zeta function regularization method
unambiguously prescribes that the sum under consideration
should be assigned the following value
\begin{equation}
\sum_{n=0}^\infty (n+\beta)=\zeta_H (-1;\beta).
\end{equation}
The wrong alternative (for obvious reasons, after all what has
been said before), would be to argue that `we might as well
have written
\begin{equation}
\sum_{n=0}^\infty (n+\beta) = \zeta_R (-1) +\beta \zeta_R (0),
\end{equation}
what gives a different result'. In fact, that `the Hurwitz
zeta function (and not the ordinary Riemann)' was the one `to
be used' was recognized by  Li et
al. [6], who reproduced in this way the correct result
obtained by Brevik and Nielsen by means of a (more
conventional)
exponential cutoff regularization. However, the authors of [6]
were again misunderstanding the main issue when they
considered their method as being a {\it generalization}  of
the zeta regularization procedure (maybe just because the
generalized Riemann zeta function appears!). Quite on the
contrary, this is just a {\it specific and
particularly simple} application of the zeta function
regularization procedure .

Of course, the method can be viewed as just one of the
many possibilities of analytic continuation  in order to
give sense to (i.e., to regularize) infinite expressions. From
this point of view, it
is very much related with the standard dimensional
regularization
method. In a very recent paper, [7], Svaiter and Svaiter have
argued that, being so close relatives, these two procedures
even share the same type of diseases. But precisely to cure
the problem of the dependence of the
regularized result on the kind of the extra dimensions
(artificially introduced in
dimensional regularization) was ---let us recall--- one of the
main motivations of Hawking for the introduction of a new
procedure, i.e. zeta function regularization, in physics [1].
So we seem to have been caught in a devil's staircase.

The solution to this paradox is the following. Actually, there
is no error in the examples of Ref. [7] and the authors know
perfectly what they are doing, but their
interpretation of the results may originate a big deal of
confusion among non-specialists.
To begin with, it might look at first sight as if the concept
itself of analytical continuation would {\it not} be {\it
uniquely}
defined. Given a function in some domain of the complex plane
(here, normally, a part of the real line or the half
plane Re $s > a$, being $a$ some abscisa of convergence), its
analytic continuation to the rest of
the complex plane (in our case, usually as a meromorphic
function, but this need not in general be so) {\it is}
uniquely
defined. Put it plain, a function {\it cannot} have two
different analytic
continuations.  What Svaiter and Svaiter do in their examples
is simply to start in each case from two different functions
of $s$
and then continue each of them analytically. Of course, the
result is different. In particular, these functions are
\begin{equation}
f_1(s) =\sum_{n=0}^\infty n^{-s}
\end{equation}
vs.
\begin{equation}
f_2(s) =\sum_{n=0}^\infty n \left(\frac{n}{a}+1
\right)^{-(s+1)}
\end{equation}
continued to $s=-1$, in the first example, which corresponds
to a Hermitian massless conformal scalar field in 2d Minkowski
spacetime with a compactified dimension, and
\begin{equation}
g_1(s) =\sum_{n=0}^\infty n^{-3s}
\end{equation}
continued to $s=-1$, vs.
\begin{equation}
g_2(s) =\sum_{n=0}^\infty n^3 \left(\frac{n}{a}+1 \right)^{-
s},
\end{equation}
continued to $s=0$, in the second example, in which the vacuum
energy corresponding to a conformally coupled scalar field
in an Einstein universe is studied. Needless to say, the
number of posibilites to define `different analytic
continuations'
in this way is literally infinite. What use can one make of
them remains to be seen.

However, what is absolutely misleading is to conclude from
those examples that analytical regularizations `suffer from
the same problem as dimensional regularization', precisely
the one that Hawking wanted to cure!. This has no meaning.
In the end, also dimensional regularization is an analytical
procedure!
One must realize that zeta function regularization is
perfectly well defined, and has little to do with these
arbitrary analytic continuations `{\it \`{a} la} $\zeta$' in
which one changes at will any exponent at any place with the
only restriction that one recovers the starting expression for
a particular value of the exponent $s$.

The facts are as follows.
(i) There exist infinitely many different analytic
regularization procedures, being dimensional regularization
and zeta function regularization just two of them. (ii) Zeta
function regularization is, as we have seen, a specifically
defined procedure, provides a unique analytical continuation
and (sometimes) a finite result. (iii) Therefore, zeta
function regularization does not suffer, in any way, from the
same kind of problem (or a related one) as dimensional
regularization. (iv) This does not mean, however, that
zeta function regularization has {\it no} problems, but they
are of a different kind; the first appears already when it
turns out that the point (let say $s=-1$ or $s=0$) at which
the zeta function must be evaluated turns out to be precisely
a {\it pole} of the analytic continuation. This and similar
difficulties can be solved, as discussed in detail in Ref.
[8]. Eventually, as a
final step one has to resort to renormalization group
techniques [9]. (v) Zeta function regularization has been
extended to higher loop order by McKeon and Sherry under the
name of operator regularization and there also some
difficulties (concerning the breaking of gauge invariance)
appear [10]. (vi) But, in the end, the
fundamental question is:  which of the regularizations that
are being used is {\it the one} choosen by nature? In
practice, one always tries to avoid
answering this question, by
cheking the finite results obtained with different
regularizations and by comparing them with classical limits
which provide well-known, physically meaningful values.
However, one would be led to believe that in view of its
uniqueness, naturalness and
mathematical elegance, zeta function regularization could well
be {\it the one}. Those properties are certainly to
be counted among their main virtues, but (oddly enough) in
some sense also as its
drawbacks: we do not manage to see clearly how and what
infinites are thrown away, something that is evident in other
more pedestrial regularizations (which are actually equivalent
in some cases to the zeta one, as pointed out, e.g., in [7]).

The final issue of this paper will concern the practical
application of the procedure. Actually, aside from some very
simple cases (among those, the ones reviewed here), the
use of the procedure of analytic continuation through
the zeta function requires a good deal of mathematical work
[2]. It is no surprise that has been so often
associated with mistakes and errors [11]. One which often
repeates itself can be traced back to Eq. (1.70) of the
celebrated book by Mostepanenko and Trunov [12] on the Casimir
effect:
\begin{equation}
\frac{a^2}{\pi^2} \sum_{n=1}^\infty  \left( n^2 +
\frac{a^2m^2}{\pi^2} \right)^{-1} =\frac{1}{2m^2} \left( -1 +
\frac{am}{\pi} \coth \frac{am}{\pi} \right),
\label{comp}
\end{equation}
in other words (for $a = \pi$ and $m=c$),
\begin{equation}
\sum_{n=1}^\infty  \left( n^2 + c^2 \right)^{-1}
=\frac{1}{2c^2} \left( -1 + c \coth c \right).
\end{equation}
That Eq. (13) is not right can be observed by simple
inspection. The corrected formula reads
\begin{equation}
\sum_{n=1}^\infty  \left( \pi^2 n^2 + c^2 \right)^{-1}
=\frac{1}{2c^2} \left( -1 + c \coth c \right).
\end{equation}

The integrated version of this equality, namely,
\begin{equation}
\sum_{n=-\infty}^\infty \ln \left( n^2 + c^2/ \pi^2 \right)
=2c+2\ln \left( 1 - e^{-2c} \right),
\end{equation}
under the specific form
\begin{equation}
T \sum_{n=-\infty}^\infty \ln \left[ ( \omega_n)^2 + (q_l)^2
\right] =q_l +2T \ln \left( 1 - e^{-q_l/T} \right),
\end{equation}
with $\omega_n=2\pi nT$ and $q_l=\pi
l/R$, has been used by Antill\'{o}n and
Germ\'{a}n in a very recent paper ([13], Eq. (2.20)), when
studying the
Nambu-Goto string model at finite length and non-zero
temperature. Now this equality is again formal. It involves an
analytic continuation, since it has no sense to integrate the
lhs term by term: we get a divergent series.

A rigorous way to proceed is as follows. The expression on the
left hand side happens to be the most
simple form of the inhomogeneous Epstein zeta function
(called usually Epstein-Hurwitz zeta function [4]).
This function is quite involved and different expresions for
it (including asymptotical expansions very useful for
accurate numerical calculations) have been given in [14] (see
also [15]). In
particular
\begin{eqnarray}
&&  \zeta_{EH}(s;c^2) =\sum_{n=1}^\infty  \left( n^2 + c^2
\right)^{-s} \nonumber \\
&& \ \ \  \  = -\frac{c^{-
2s}}{2} + \frac{\sqrt{\pi} \, \Gamma (s- 1/2)}{2\Gamma (s)}
c^{-2s+1} + \frac{2\pi^s c^{-s +1/2}}{\Gamma (s)}
\sum_{n=1}^\infty n^{s -1/2} K_{s -1/2} (2\pi nc),
\label{my}
\end{eqnarray}
which is reminiscent of the famous Chowla-Selberg
formula (see [3], p. 1379). Derivatives can be taken here and
the analytical continuation in $s$ presents again no problem.

The usefulness of zeta function regularization is without
question [16,2,4]. It can give immediate sense to expressions
such as $1+1+1+\cdots =-1/2$, which turn out to be invaluable
for the construction of new physical theories, as different as
Pauli-Villars regularization with infinite constants
(advocated by Slavnov [17]) and mass generation in cosmology
through Landau poles (used by Yndurain [18]). The Riemann zeta
function was termed by Hilbert in his famous 1900 lecture as
the most important function of whole mathematics [19].
Probably it will remain so in the Paris Congress of AD
2000, but now maybe with quantum field physics adhered to.
\vspace{5mm}

\noindent{\large \bf Acknowledgments}

I am very grateful to Prof.~I.~Brevik  and  to
Prof.~K.~Olaussen for many illuminating
discussions and also to them and to Prof.~L.~Brink for the
hospitality extended to me at
the Universities of Trondheim and G\"{o}teborg, respectively.
This work has been  supported by DGICYT (Spain) and by CIRIT
(Generalitat de Catalunya).

\end{document}